\documentclass[twocolumn,showpacs,preprintnumbers,amsmath,amssymb]{revtex4}
\usepackage{graphicx}
\usepackage{dcolumn}
\usepackage{bm}
\begin{document}
\title{Fast exciton spin relaxation in single quantum dots}
\author{I. Favero,$^1$ G. Cassabois,$^{1,\ast}$ C. Voisin,$^1$ C. Delalande,$^1$
 Ph. Roussignol,$^1$ R. Ferreira,$^1$ C. Couteau,$^2$ J. P. Poizat,$^2$ and J. M. G\'{e}rard$^3$}
\affiliation{$^1$Laboratoire Pierre Aigrain, Ecole Normale
Sup\'erieure, 24 rue Lhomond 75231 Paris Cedex 5, France\\
$^2$CEA-CNRS-UJF "Nanophysics and Semiconductors" Laboratory,
Laboratoire de Spectrom\'{e}trie Physique, 140 Avenue de la
Physique - BP 87, 38402 Saint Martin d'H\`{e}res, France\\
$^3$CEA-CNRS-UJF "Nanophysics and Semiconductors" Laboratory,
CEA/DRFMC/SP2M, 17 rue des Martyrs 38054 Grenoble Cedex 9, France}
\date{\today}
\begin{abstract}
Exciton spin relaxation is investigated in single epitaxially
grown semiconductor quantum dots in order to test the expected
spin relaxation quenching in this system. We study the
polarization anisotropy of the photoluminescence signal emitted by
isolated quantum dots under steady-state or pulsed non-resonant
excitation. We find that the longitudinal exciton spin relaxation
time is strikingly short ($\leq$100 ps) even at low temperature.
This result breaks down the picture of a frozen exciton spin in
quantum dots.
\end{abstract}

\maketitle

Spin memory effects in semiconductor quantum dots (QDs) attract
presently much attention in the physics of nanostructures. The
discrete energy spectrum of zero-dimensional carriers in QDs is
expected to lead to an inhibition of the main spin relaxation
mechanisms which are known in bulk semiconductors and planar
heterostructures \cite{meier,maialle,khaetskii}. In some novel QD
devices, the preservation of the exciton spin coherence is a
central issue, for instance for the generation of
polarization-entangled photon-pairs in quantum information
processing \cite{moreau,santori}. Recent studies of epitaxially
grown InGaAs/GaAs QDs have shown that the longitudinal exciton
spin relaxation may be quenched over tens of ns at low temperature
\cite{paillard,lenihan,langbein}. However, it was also suggested
in Ref.~\cite{lenihan} that some QDs could undergo a rapid spin
relaxation, because the longitudinal spin dynamics exhibits, for
QD arrays, a fast decay component (40 ps). This fact highlights
the need for experiments probing spin relaxation dynamics on the
single QD level. However, the implementation of the standard
time-resolved techniques used for QDs ensembles
\cite{paillard,lenihan,langbein} still remains an experimental
challenge in the field of single QD spectroscopy.

In this letter, we report on the observation of a strikingly fast
longitudinal relaxation ($\leq$100 ps) of the exciton spin for
some single InAs/GaAs QDs. We study single neutral QDs, isolated
from a dilute QD array by micro-photoluminescence (PL). We focus
our attention on QDs displaying a polarization anisotropy of their
emission under non-resonant excitation. On the basis of
time-resolved data and rate equation analysis, we demonstrate that
this feature can only be understood as resulting from a fast
direct transfer between the bright exciton states, leading to spin
relaxation. This result breaks down the picture of an universal
freezing of the exciton spin in QDs at low temperature.

Let us first describe the fine structure of the exciton ground
state in a neutral QD, and in particular the polarization
anisotropy underlying our experimental approach. Assuming
rotational symmetry of the QD along the [001] growth axis, the
ground heavy-hole exciton state is fourfold degenerate with
electron-hole pair states having a projection of the total angular
momentum along the growth axis $J_z$ equal to $\pm$1 and $\pm$2.
In the dipole approximation, the two $|$$J_z$=$\pm$2$\rangle$
states are dark and the two $|$$J_z$=$\pm$1$\rangle$ states are
bright and coupled to orthogonal circularly polarized light
states. The short-range part of the electron-hole exchange
interaction lifts their degeneracy and the dark states lie a few
hundreds of $\mu$eV below the degenerate bright states
\cite{bayer}. However, different microscopic effects such as the
anisotropy of the QD confinement potential but also the atomistic
symmetry of the crystal \cite{zunger} modify this simple picture
and break the QD rotational invariance. The fine structure of the
exciton ground state is determined by the subtle interplay between
the spin-orbit and exchange interactions as a function of the
crystal atomistic symmetry and QD shape, as discussed in detail in
Ref.~\cite{zunger}. The most important result is that the new
bright states $|X\rangle$ and $|Y\rangle$ are no longer degenerate
and correspond to two linearly polarized transitions which are
aligned along two orthogonal principal axes of the QD.
Furthermore, the symmetry reduction gives rise, through a
heavy-light hole mixing \cite{tanaka} or through a deformation of
the wave-function envelope, to different oscillator strengths
$\gamma_X$ and $\gamma_Y$ for the bright states \cite{wang}. This
phenomenon is the origin of the polarization anisotropy of the PL
emission.

The observation of split linearly polarized transitions has been
reported for QDs of various materials by means of single-QD
spectroscopy \cite{gammon,bayer2,besombes,santori}. The ratio of
the PL intensities of the $|X\rangle$ and $|Y\rangle$ states
varies from one study to another and the observed polarization
anisotropy is naturally interpreted in terms of different
oscillator strengths for the two bright exciton states in a QD of
reduced symmetry. However, to the best of our knowledge, it has
never been pointed out that the observation of polarization
anisotropy in PL measurements is not possible if the dynamics of
the system is only driven by radiative recombination. Another
relaxation mechanism is needed in order to obtain different PL
intensities for the $|X\rangle$ and $|Y\rangle$ lines, as
explained below.

\begin{figure}
\includegraphics[scale=0.35,angle=-90]{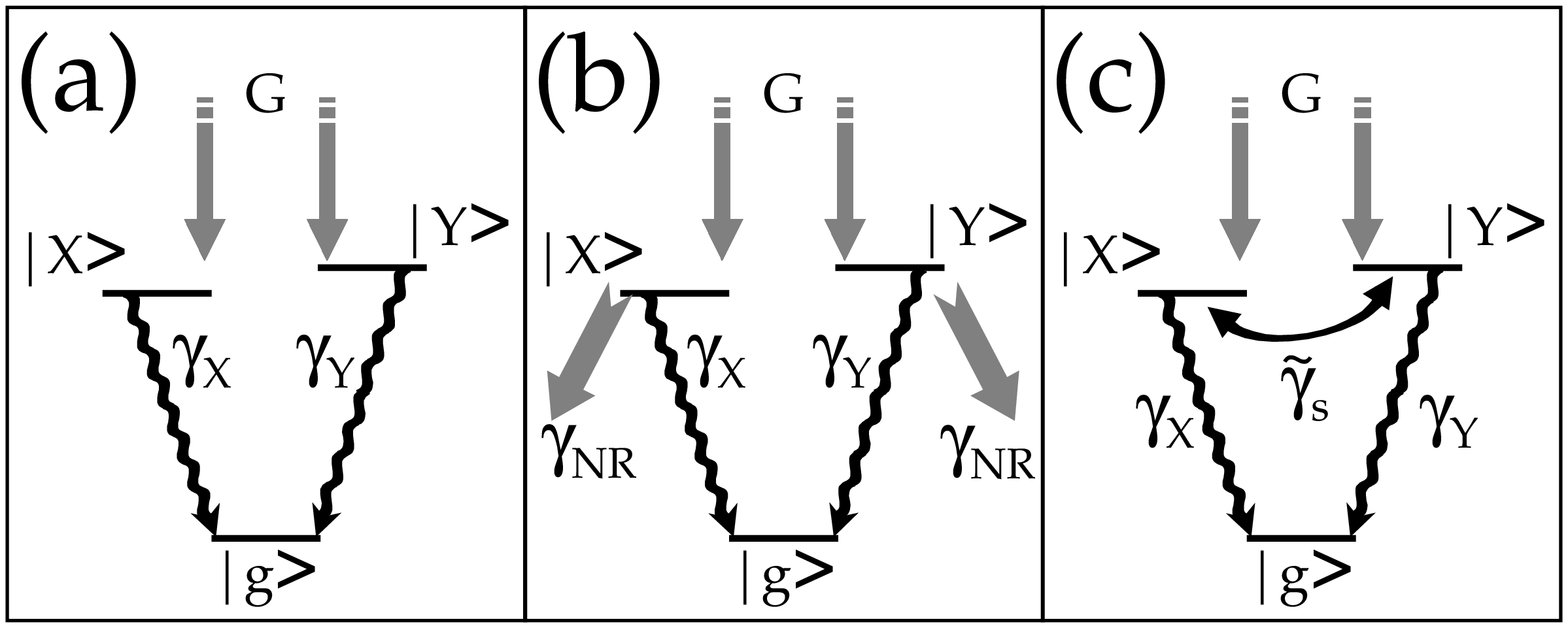}
\caption{Energy level diagram of a QD showing the two bright
exciton states $|X\rangle$ and $|Y\rangle$ and the system ground
state $|g\rangle$. (a) The system dynamics is determined by an
exciton photo-generation rate $G$ and two oscillator strengths
$\gamma_X$ and $\gamma_Y$ accounting for radiative recombination.
An additional relaxation channel is considered : non-radiative
recombination (b) or longitudinal exciton spin relaxation (c).}
\label{RateEq}
\end{figure}

For the sake of clarity, we first analyze a three-level system
with the two bright exciton states $|X\rangle$ and $|Y\rangle$ and
the ground state $|g\rangle$ [Fig.~\ref{RateEq}]. We examine the
results of a simple rate equation model where the QD states are
equally populated with a photo-generation rate $G$ and radiatively
recombine with respective oscillator strengths $\gamma_X$ and
$\gamma_Y$. The fact that both bright states are equally populated
is a central assumption of our model and it requires a
non-resonant excitation of the QD to avoid any polarization memory
of the excitation laser, as detailed below. If the system dynamics
is only driven by radiative recombination [Fig.~\ref{RateEq}(a)],
the steady-state PL intensity of both bright exciton states in the
weak excitation regime is \textit{exactly the same} whatever the
difference in oscillator strengths. This result may be surprising
at first sight but simply stems from a quantum efficiency equal to
unity for both $|X\rangle$ and $|Y\rangle$ states: every single
exciton which is photo-generated on a bright state will give rise
to the emission of a single photon. While the steady-state
populations $n_X$ and $n_Y$ do depend on the oscillator strengths
$\gamma_X$ and $\gamma_Y$ ($n_\beta$=$G$/$\gamma_\beta$ with
$\beta$=$X$ or $Y$), the PL intensities are only proportional to
the pumping rate
($I_{|\beta\rangle}$$\propto$$\gamma_\beta$$n_\beta$$\propto$$G$).

This means that the polarization anisotropy observed for the QD
exciton ground state in PL experiments has to be revealed by an
additional relaxation channel. In the following we consider two
possible schemes for this additional process: non-radiative
recombination [Fig.~\ref{RateEq}(b)] and longitudinal exciton spin
relaxation [Fig.~\ref{RateEq}(c)]. In both cases the PL signal
exhibits a polarization anisotropy with a linear polarization
ratio $R_L$ given by:
\begin{equation}
  R_L=\frac{I_{|X\rangle}-I_{|Y\rangle}}{I_{|X\rangle}+I_{|Y\rangle}}=\frac{\gamma_X-\gamma_Y}{\gamma_X+\gamma_Y+\gamma_X\gamma_Y/\Gamma}
\label{Pratio}
\end{equation}
where $\Gamma$ is equal to $\gamma_{NR}/2$ and
$\widetilde{\gamma_{s}}$ for cases (b) and (c) of
Fig.~\ref{RateEq}, respectively. In the limit of a small
perturbation to the radiative recombination
($\Gamma$$\ll$$\gamma_X$,$\gamma_Y$) the polarization anisotropy
vanishes ($R_L$$\sim$0), as explained above. On the contrary, when
the additional relaxation channel strongly competes with the
radiative coupling ($\Gamma$$\gg$$\gamma_X$,$\gamma_Y$) the linear
polarization ratio $R_L$ is equal to
($\gamma_X$-$\gamma_Y$)/($\gamma_X$+$\gamma_Y$) revealing the
different oscillator strengths of the bright exciton states. In
fact, an efficient non-radiative recombination or spin relaxation
equalizes the populations of the two bright states so that the PL
intensities become proportional to their respective oscillator
strengths
($I_{|\beta\rangle}$$\propto$$\gamma_\beta$$\widetilde{n}$ where
$\widetilde{n}$ is the common value of $n_X$ and $n_Y$). In the
following we present polarization-resolved PL experiments in
single InAs/GaAs QDs which provide a clear insight into the
exciton spin relaxation dynamics in isolated QDs.

Single-QD spectroscopy at low temperature is performed by standard
micro-PL measurements in the far field using a microscope
objective in a confocal geometry. The excitation beam is provided
by a He:Ne laser and a frequency-doubled cw mode-locked Ti:Sa
laser for steady-state and time-resolved PL experiments,
respectively. In the steady-state PL measurements, the signal is
detected, after spectral filtering by a 32 cm monochromator, by a
LN$_{2}$-cooled charge-coupled-device (CCD) with a spectral
resolution of 120 $\mu$eV. Life-time measurements are performed by
time-correlation single photon counting with a temporal resolution
of 400 ps. Polarization-resolved experiments are implemented by
using a half-wave retarder and a fixed linear polarizer in front
of the spectrometer in order to avoid any detection artifacts due
to the anisotropic response function of the setup.

\begin{figure}
\begin{center}
\includegraphics[scale=0.95]{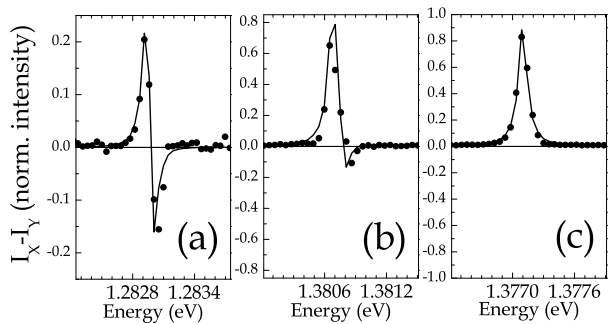}
\caption{Normalized differential photoluminescence spectra
$I_X$-$I_Y$ for three quantum dots at 10K, where $I_X$ and $I_Y$
are the photoluminescence spectra recorded for the $X$ and $Y$
analysis axes, respectively. Fits (solid line) are performed with
two orthogonal linearly polarized lines with
$I_{|X\rangle}$/$I_{|Y\rangle}$ equal to 1.2 (a), 3 (b), and 10
(c), and an energy splitting $\Delta$$\sim$35 $\mu$eV (a),
$\Delta$$\sim$105 $\mu$eV (b), and $\Delta$$<$40 $\mu$eV (c).}
\label{Disp}
\end{center}
\end{figure}

In Fig.~\ref{Disp} we display the polarization-resolved PL
measurements under steady-state excitation, at low temperature for
three types of QDs in a dilute QD array \cite{favero}. In order to
better visualize the fine structure of the exciton ground state,
we plot differential PL spectra $I_X$-$I_Y$ where we subtract the
PL spectra recorded for the $X$ and $Y$ analysis axes,
corresponding to the [110] and [1$\overline{1}$0] crystallographic
directions. In the dispersive profiles, the positive signal
corresponds to the emission of the $|X\rangle$ state and the
negative one to the $|Y\rangle$ state. After normalization by the
PL intensity maximum of the total spectrum $I_X$+$I_Y$, we
carefully fit our dispersive profiles with theoretical curves
calculated with two orthogonal linearly polarized lines of
different intensities and split by an energy $\Delta$. Such a
procedure allows us to resolve an energy splitting below the
spectral resolution of our setup as shown in Fig.~\ref{Disp}(a)
($\Delta$$\sim$35 $\mu$eV) and (b) ($\Delta$$\sim$105 $\mu$eV). We
are nevertheless limited by the spectral extension of 45 $\mu$eV
corresponding to a single CCD pixel, and also by the PL intensity
asymmetry $I_{|X\rangle}$:$I_{|Y\rangle}$ as shown in
Fig.~\ref{Disp}(c) for a QD with
$I_{|X\rangle}$/$I_{|Y\rangle}$$\sim$10 where we only get an upper
value of 40 $\mu$eV for the bright states energy splitting.

Before analyzing the dynamics of the population transfer between
the bright and dark states, we would like to discuss the central
assumption of our model where we consider the same exciton
photo-generation rate $G$ for both bright states. Indeed,
different photo-generation rates $G_X$ and $G_Y$ for two bright
exciton states of identical oscillator strengths would also lead
to a polarization anisotropy. In fact, in both steady-state and
time-resolved PL measurements, we perform a highly non-resonant
excitation of the QDs where electron-hole pair states are
photo-created in the GaAs barrier and we do not measure any
modification of the linear polarization ratio $R_L$ after rotation
of the linear polarization of the excitation laser. This
observation indicates that the memory of the laser polarization is
lost in our experimental configuration and does not bias the
carrier relaxation into the QDs. Furthermore, we rule out a
differential exciton capture because, if $G_X$$\neq$$G_Y$, $R_L$
is predicted to decrease above the QD saturation threshold in a
model including multi-excitonic states \cite{dekel}, whereas in
our power-dependent measurements $R_L$ remains constant. We
therefore validate the assumption of an identical exciton
photo-generation rate $G$ for both bright states and we now
comment on our time-resolved PL experiments.

\begin{figure}
\begin{center}
\includegraphics[scale=0.75]{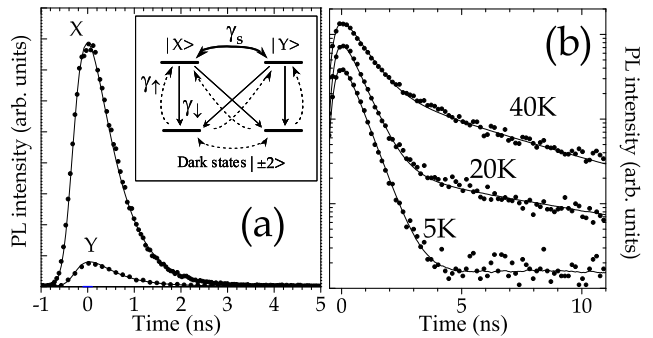}
\caption{(a) Time-resolved photoluminescence intensity at 5K, for
the $X$ (top curve) and $Y$ (bottom curve) polarization analysis
axes, for the QD of Fig.~\ref{Disp}(c). (Inset): Schematic diagram
of the exciton ground state fine structure. (b) Time-resolved
traces for the $X$ analysis axis at 5, 20 and 40K on a
semi-logarithmic scale. Fits (solid line) are solutions of the
rate equation model sketched in the inset.} \label{TR}
\end{center}
\end{figure}

The PL decay dynamics investigated for a set of ten QDs show
qualitatively the same features: (i) a bi-exponential decay of the
population dynamics when the temperature increases
[Fig.~\ref{TR}(b)] which is the signature of a coupling to the
dark states \cite{lounis}, (ii) a constant ratio
$I_{|X\rangle}(t)$/$I_{|Y\rangle}(t)$ of the time-resolved PL
intensities of the bright states [Fig.~\ref{TR}(a)] whatever the
temperature. Note that this latter fact implies that the
time-resolved linear polarization ratio $R_L(t)$ does not exhibit
any variation as a function of time.

In Fig.~\ref{TR}(a) we display typical time-resolved PL
measurements at 5K for the $X$ (top curve) and $Y$ (bottom curve)
polarization analysis axes. Both curves are well fitted by the
numerical convolution of the temporal response function of the
setup with an exponential of 450 ps time constant. The ratio
$I_{|X\rangle}(t)$/$I_{|Y\rangle}(t)$ has thus a constant value
which is \textit{moreover the same as in steady-state
measurements}. Such a result is the counterpart in the temporal
domain of the phenomenology discussed above in stationary
conditions. Namely, in the regime where the additional relaxation
channel strongly competes with the radiative coupling
($\Gamma$$\gg$$\gamma_X$,$\gamma_Y$), the transient populations of
the bright exciton states rapidly equalize so that we monitor the
global decay of the bright exciton states of different oscillator
strengths
($I_{|\beta\rangle}(t)$$\propto$$\gamma_\beta$$\widetilde{n}(t)$
where $\widetilde{n}(t)$ is the common value reached by the
transient exciton populations).

Let us now discuss the additional relaxation process.
Non-radiative recombination is known to be negligible in InAs QDs
at low temperature \cite{gerard,robert}. Moreover, its efficiency
is expected to increase with temperature so that the presence of
non-radiative recombination centers would hasten the PL decay when
rising the temperature. On the contrary, an overall slow-down of
the carrier dynamics is observed in our experiments
[Fig.~\ref{TR}(b)] with a bi-exponential decay of the PL dynamics
due to the exciton dark states \cite{lounis}. We therefore
conclude that non-radiative recombination does not play a major
role in the bright states recombination [Fig.~\ref{RateEq}(b)] and
that a fast exciton spin relaxation occurs in our single QDs
[Fig.~\ref{RateEq}(c)].

In order to get quantitative information on the exciton spin
relaxation, we generalize the model of Fig.~\ref{RateEq}(c) by
taking into account the dark exciton states which lead to the
bi-exponential decay of the population dynamics
[Fig.~\ref{TR}(b)]. Extending the approach of Labeau \textit{et
al.} \cite{lounis}, we use a five-level system and we calculate
the population dynamics according to the model sketched in the
inset of Fig.~\ref{TR}. The population relaxation rate from a
bright to a dark state is $\gamma_0$ at zero temperature. We
account for an acoustic phonon-assisted thermalization with a rate
$\gamma_\downarrow=\gamma_0(N+1)$ for each bright-to-dark channel
and $\gamma_\uparrow=\gamma_0N$ for each dark-to-bright one, where
$N=1/[\exp(\Omega/k_BT)-1]$ is the Bose-Einstein phonon number and
$\Omega$ is the energy splitting between the bright and dark
states, which lifetime is 1/$\gamma_D$. For the population
relaxation among the bright states, we take the same transfer rate
$\gamma_s$ for both paths because $\Delta$$\ll$$k_BT$. This model
yields, for the steady-state linear polarization ratio $R_L$, the
expression given in Eq.~\ref{Pratio} where the effective spin
relaxation rate $\widetilde{\gamma_{s}}$ introduced in
Fig.~\ref{RateEq}(c) is equal to $\gamma_s+\gamma_\downarrow$. As
far as the time-resolved experiments are concerned, we assume that
the bright and dark states are equally populated at $t$=0 and we
plot in solid lines in Fig.~\ref{TR}(a) and (b) the solutions of
our rate equation model after numerical convolution with the
temporal response function of the setup. We observe a fair
agreement with all our time-resolved measurements, for realistic
values of 1/$\gamma_D$$\sim$8.5 ns and $\Omega$$\sim$250 $\mu$eV
\cite{bayer}.

The efficient direct population transfer between the $|X\rangle$
and $|Y\rangle$ states is the origin of the constant ratio
$I_{|X\rangle}(t)$/$I_{|Y\rangle}(t)$ of the time-resolved PL
intensities [Fig.~\ref{TR}(a)], and we find that the corresponding
time constant is smaller than 100 ps (1/$\gamma_s\leq$0.1 ns). At
low temperature (5K) the influence of the dark states on the
exciton dynamics is only marginal. The quasi-exponential dynamics
of the global population of the bright states has a decay time
$\tau$ corresponding to the oscillator strengths average
(1/$\tau$$\sim$($\gamma_X$+$\gamma_Y$)/2). Given the oscillator
strengths ratio
$\gamma_X$/$\gamma_Y$$\sim$$I_{|X\rangle}(t)$/$I_{|Y\rangle}(t)$$\sim$10,
we obtain radiative lifetimes 1/$\gamma_X$$\sim$0.24 ns and
1/$\gamma_Y$$\sim$2.4 ns. Indeed, since spin relaxation is faster
than radiative recombination, we do not observe the separate
decays of the bright states populations but their global common
relaxation. If we decrease the population transfer rate $\gamma_s$
in our simulations, we obtain distinct dynamics for the transient
populations $n_X(t)$ and $n_Y(t)$ as well as a decrease of the
steady-state value of the linear polarization ratio $R_L$, in
contradiction to our experimental findings. Such an efficient
exciton spin relaxation is a very important and unexpected
conclusion of our study that applies to all QDs showing any
detectable polarization anisotropy in their PL emission under
non-resonant excitation. Complementary measurements (to be
reported elsewhere) show that the breakdown of the frozen exciton
spin picture can also be observed in QDs ensembles with a weak
polarization anisotropy. The striking fast exciton spin relaxation
in neutral QDs raises fundamental theoretical questions on the
corresponding microscopic mechanisms, which are beyond the scope
of this work.

We finally discuss whether the indirect channel opened by the
coupling to the dark states can also result in an efficient
population transfer between the bright states. The relaxation
rates to the dark states are determined from the analysis of the
temperature-dependent data [Fig.~\ref{TR}(b)]. When the
temperature increases, the gradual thermalization of the bright
and dark states populations gives rise to the bi-exponential decay
where the slow component decay time is related to the dark state
lifetime. Note that a complete thermalization would lead to a
quasi-exponential PL decay determined by the global dynamics of
the bright and dark states. From our temperature-dependent data,
we deduce a zero-temperature relaxation time 1/$\gamma_0$$\sim$440
ns. For this QD, the relaxation to the dark states plays a minor
role compared to the direct population transfer between bright
excitons. On the other hand, for another QD (not shown) we have
measured a zero-temperature relaxation time as small as
1/$\gamma_0$$\sim$30 ns which is one order of magnitude shorter
than above. The corresponding effective spin relaxation is
$1/\gamma_\downarrow$$\sim$13 ns at 5K and 2 ns at 40K. This
results demonstrates that the population redistribution via the
dark states remains anyway negligible for this QD compared to the
direct transfer between the bright states
($\gamma_\downarrow$$\ll$$\gamma_s$ in
$\widetilde{\gamma_s}$=$\gamma_s$+$\gamma_\downarrow$).

In summary, we have studied the longitudinal exciton spin
relaxation dynamics in single QDs by analyzing the population
transfer between the bright and dark states forming the exciton
fine structure in neutral InAs quantum dots. We find that the
longitudinal exciton spin relaxation time is strikingly short
($\leq$100 ps) even at low temperature. This result breaks down
the picture of an universal freezing of the exciton spin for QDs
at low temperature. Further work is clearly needed to elucidate
the interplay between the fine structure and the exciton spin
relaxation, and to determine the experimental parameters that
control the efficiency of spin relaxation.

LPA-ENS is "unit\'{e} mixte (UMR 8551) de l'ENS, du CNRS, des
Universit\'{e}s Paris 6 et 7". This work is financially supported
by the ACI "Polqua", and by the region Ile de France through the
project SESAME E-1751.

$^\ast$Electronic address: Guillaume.Cassabois@lpa.ens.fr

\end{document}